\documentstyle[12pt,epic]{article}
\renewcommand{\title}[1]{\large\bf 
     #1\bigskip\medskip\\} 
\renewcommand{\author}[1]{\large #1\\ \smallskip}
\newcommand{\address}[1]{{\normalsize\it #1\\}\bigskip}

\newcommand{\be}{\begin{eqnarray}}
\newcommand{\ee}{\end{eqnarray}} 
\newcommand{\hs}[1]{\hspace*{#1cm}}

\newcommand{\no}{\nonumber}
\newcommand{\ig}[1]{\mbox{}}
\def\trans#1#2#3#4#5#6#7#8#9{\begin{picture}(320,85)(80,715)
\multiput(120,760)(80,0){4}{\circle*{6}}
\multiput(160,720)(80,0){3}{\circle*{#5}}
\put( 80,800){\line( 1,-1){ 80}}
\put(160,720){\line( 1, 1){ 80}}
\put(240,800){\line( 1,-1){ 80}}
\put(320,720){\line( 1, 1){ 80}}
\put( 80,720){\line( 1, 1){ 80}}
\put(160,800){\line( 1,-1){ 80}}
\put(240,720){\line( 1, 1){ 80}}
\put(320,800){\line( 1,-1){ 80}}
\put(180,775){\tiny$#1$}
\put(180,740){\tiny$#2$}
\put(135,735){\tiny$#4$}
\put(135,775){\tiny$#3$}
\put(70,800){\tiny$#6$}
\put(70,720){\tiny$#7$}
\put(400,800){\tiny$#8$}
\put(400,720){\tiny$#9$}
\end{picture}}

\def\bottom{
\begin{picture}(350,12)(70,695)
\multiput(240,700)(8.63636,0.00000){3}{\makebox(0.4444,0.6667){.}}
\put(120,695){\makebox(0,0)[lb]{\tiny$$}}
\put(160,695){\makebox(0,0)[lb]{\tiny$2$}}
\put(390,695){\makebox(0,0)[lb]{\tiny$L\!+\!1$}}
\put( 80,695){\makebox(0,0)[lb]{\tiny$1$}}
\end{picture}}

\def\bond#1#2#3#4{\setlength{\unitlength}{0.00800000in}%
\begin{picture}(6,72)(123,714)
\multiput(120,780)(0.25000,-0.50000){13}{.}
\multiput(123,774)(-0.17647,-0.52941){18}{.}
\multiput(120,765)(-0.25000,-0.50000){13}{.}
\multiput(117,759)(0.17647,-0.52941){18}{.}
\multiput(120,750)(0.25000,-0.50000){13}{.}
\multiput(123,744)(-0.17647,-0.52941){18}{.}
\multiput(120,735)(-0.25000,-0.50000){13}{.}
\multiput(117,729)(0.17647,-0.52941){18}{.}
\multiput(122,783)(0,-64){2}{\circle*{#4}}
\put(110,779){\makebox(0,0)[lb]{\small$#1$}}
\put(111,710){\makebox(0,0)[lb]{\small$#2$}}
\put(125,741){\makebox(0,0)[lb]{\tiny$#3$}}
\end{picture}}

\def\face#1#2#3#4{\setlength{\unitlength}{0.00800000in}%
\begin{picture}(6,69)(114,699)
\put(120,765){\line( 0,-1){ 60}}
\put(110,762){\small$#1$}
\put(110,699){\small$#2$}
\put(114,726){\tiny$#3$}
\multiput(120,765)(0,-60){2}{\circle*{#4}}
\end{picture}}

\def\st#1#2#3#4#5#6#7{\setlength{\unitlength}{0.0067000in}%
\begin{picture}(170,117)(99,678)
\put(150,735){\circle*{6}}
\multiput(150,780)(0.25000,-0.50000){13}{.}
\multiput(153,774)(-0.17647,-0.52941){18}{.}
\multiput(150,765)(-0.25000,-0.50000){13}{.}
\multiput(147,759)(0.17647,-0.52941){18}{.}
\multiput(150,750)(0.25000,-0.50000){13}{.}
\multiput(153,744)(-0.17647,-0.52941){18}{.}
\multiput(150,735)(-0.17647,-0.52941){18}{.}
\multiput(147,726)(0.25000,-0.50000){13}{.}
\multiput(150,720)(0.25000,-0.50000){13}{.}
\multiput(153,714)(-0.17647,-0.52941){18}{.}
\multiput(150,705)(-0.17647,-0.52941){18}{.}
\multiput(147,696)(0.25000,-0.50000){13}{.}
\put(105,735){\line( 1, 0){ 45}}
\put(105,780){\line( 0,-1){ 45}}
\put(153,753){\makebox(0,0)[lb]{\tiny$#6$}}
\put(153,699){\makebox(0,0)[lb]{\tiny$#7$}}
\put(108,753){\makebox(0,0)[lb]{\tiny$#4$}}
\put(126,738){\makebox(0,0)[lb]{\tiny$#5$}}
\put( 99,780){\makebox(0,0)[lb]{\small$#1$}}
\put( 99,729){\makebox(0,0)[lb]{\small$#2$}}
\put(144,783){\makebox(0,0)[lb]{\small$#1$}}
\put(144,681){\makebox(0,0)[lb]{\small$#3$}}
\put(264,732){\circle*{6}}
\multiput(264,780)(0.25000,-0.50000){13}{.}
\multiput(267,774)(-0.17647,-0.52941){18}{.}
\multiput(264,765)(-0.25000,-0.50000){13}{.}
\multiput(261,759)(0.17647,-0.52941){18}{.}
\multiput(264,750)(0.25000,-0.50000){13}{.}
\multiput(267,744)(-0.17647,-0.52941){18}{.}
\multiput(264,735)(-0.17647,-0.52941){18}{.}
\multiput(261,726)(0.25000,-0.50000){13}{.}
\multiput(264,720)(0.25000,-0.50000){13}{.}
\multiput(267,714)(-0.17647,-0.52941){18}{.}
\multiput(264,705)(-0.17647,-0.52941){18}{.}
\multiput(261,696)(0.25000,-0.50000){13}{.}
\put(219,687){\line( 0, 1){ 45}}
\put(219,732){\line( 1, 0){ 45}}
\put(261,780){\makebox(0,0)[lb]{\small$#1$}}
\put(210,729){\makebox(0,0)[lb]{\small$#2$}}
\put(210,684){\makebox(0,0)[lb]{\small$#3$}}
\put(255,684){\makebox(0,0)[lb]{\small$#3$}}
\put(222,702){\makebox(0,0)[lb]{\tiny$#4$}}
\put(237,735){\makebox(0,0)[lb]{\tiny$#5$}}
\put(267,753){\makebox(0,0)[lb]{\tiny$#7$}}
\put(267,693){\makebox(0,0)[lb]{\tiny$#6$}}
\put(180,732){\makebox(0,0)[lb]{$=$}}
\end{picture}}

\def\star#1#2#3{
\setlength{\unitlength}{0.01150000in}%
\begin{picture}(145,65)(75,750)
\put(210,780){\circle*{5}}
\put(100,800){\line( 0,-1){ 40}}
\put(100,760){\line( 2, 1){ 40}}
\put(140,780){\line(-2, 1){ 40}}
\put(180,800){\line( 3,-2){ 30}}
\put(210,780){\line(-3,-2){ 30}}
\put(210,780){\line( 1, 0){ 30}}
\put( 95,756){\small$c$}
\put(100,803){\small$a$}
\put(140,775){\small$b$}
\put(175,803){\small$a$}
\put(175,756){\small$c$}
\put(90,775){\tiny${\overline{W}(#2)}$}
\put(120,765){\tiny${W(#3)}$}
\put(120,790){\tiny${W(#1)}$}
\put(210,782){\tiny$W(#2)$}
\put(190,762){\tiny$\overline{W}(#1)$}
\put(190,793){\tiny$\overline{W}(#3)$}
\put(240,775){\small$b$}
\put(151,775){$=\chi$}
\end{picture}}

\def\bwa#1#2#3#4{
\begin{picture}(30,23)(69,741)
#4{\multiput(75,750)(2.5,0){13}{.}\multiput(75,750)(0,2.5){13}{.}
\multiput(105,780)(0,-2.5){13}{.}\multiput(105,780)(-2.5,0){13}{.}
\multiput(75,759)(0.4,-0.4){23}{.}
\thicklines}\put(106.5,780){\line(-1,-1){29}}
\put(70,745){\small$#1$}
\put(108,779){\small$#2$}\put( 81,763){\tiny$#3$}
\end{picture}}

\def\wb#1#2#3#4{
\begin{picture}(30,23)(69,741)
#4{\multiput(75,750)(2.5,0){13}{.}\multiput(75,750)(0,2.5){13}{.}
\multiput(105,780)(0,-2.5){13}{.}\multiput(105,780)(-2.5,0){13}{.}
\multiput(75,759)(0.4,-0.4){23}{.}\thicklines}
\put(76.9,780){\line(1,-1){29}}
\put(106,745){\small$#1$}
\put(69,779){\small$#2$}\put( 81,763){\tiny$#3$}
\end{picture}}

\begin{document}

\begin{center}
\title{ Fateev-Zamolodchikov and Kashiwara-Miwa models: \\
 boundary star-triangle relations and  surface critical properties  }
\author{Yu-Kui Zhou\footnote{ on leave from Northwest University, China.} }
\address{Department of Mathematics, School of Mathematical Sciences,\\
         The Australian National University, Canberra ACT 0200, Australia}

\begin{abstract}
The  boundary  Boltzmann weights are found by solving the boundary 
star-triangle relations for the Fateev-Zamolodchikov and
Kashiwara-Miwa models. We calculate the surface free energies of the 
models. The critical surface exponent $\alpha_s$ of the 
Kashiwara-Miwa model is given and satisfies the scaling relation 
$\alpha_b=2\alpha_s-2$, where $\alpha_b$ is the bulk exponent. 
\end{abstract}
\end{center}

\begin{quote}
 
\noindent
key-words:\newline
exactly solved models; 
boundary Boltzmann weights; boundary star-triangle relations;
surface critical phenomena; surface free energies.
\end{quote} 
\vskip 1.5cm
\subsection{Introduction}
Recently much research interest has been attracted to study exactly solvable 
models or integrable models with  boundaries
\cite{GhZa:94,JKKKM:95,MeNe:95,Saleur,Affleck}. 
In statistical mechanics it has been shown that the non-periodic 
integrable lattice models with reflection boundaries are useful 
to exploit surface critical properties  
(for reviews see \cite{Zhou:96c,Murray:96} and references therein).

In the study of the six-vertex model \cite{Sklyanin}  it has been made 
clear that the integrability of reflection boundary models
is governed by the Yang-Baxter equation \cite{Yang:67,Baxter:82} of 
the bulk Boltzmann weights and the boundary Yang-Baxter equation (or 
reflection equation \cite{Cherednik}) of the boundary Boltzmann 
weights. The boundary Boltzmann weights have been calculated for
many exactly solvable models \cite{MeNe:91}-\cite{BFKZ:96}. 
The surface properties of these models
deserve to be investigated (\cite{BFKSZ:96,OBPe:96}).  

The  Fateev-Zamolodchikov model \cite{FaZa:82}, Kashiwara-Miwa 
model \cite{KaMi:86} and chiral Potts model 
\cite{AMPTY:87,BPA:88} are very interesting $Z_N$ models
in statistical mechanics. Of these models the bulk Boltzmann weights  
satisfy the star-triangle relations (see equ.(\ref{star})), the simplest
form of the Yang-Baxter equation as stated in \cite{Baxter:96}.
The integrable boundary weights and the corresponding surface
behaviour of the $Z_N$-models have been not studied. 
In this paper the Fateev-Zamolodchikov and Kashiwara-Miwa models are
considered. We construct the boundary Boltzmann weights and 
their  boundary star-triangle relations,  the simplest form
of the boundary Yang-Baxter equation (or reflection equations).
The corresponding surface free energies are calculated. The excess
surface critical exponent of the Kashiwara-Miwa model is given. 
The chiral Potts model is not studied in this paper. This model
will be considered separately because of the absence of spectral 
difference property.

The layout of this paper is as follows. In the next  three sections a 
general consideration is treated. This means that the results given are
applicable for  the Fateev-Zamolodchikov model and Kashiwara-Miwa
model, and for other $Z_N$ models provided their bulk Boltzmann 
weights satisfy  some required crossing and inversion relations. 
In section~2 we define the double-row transfer matrix with the left
and right boundaries. Imposing the commutativity of the transfer matrix with
different spectral parameters, the boundary star-triangle relations
are extracted. The left and right boundary Boltzmann weights are determined by
the boundary star-triangle relations. In section~3 we discuss the fusion 
procedure of the $Z_N$ models briefly. Then the functional relations
of the fused transfer matrices are given. In section~4 we show  the
spectral-independent\footnote{spectral parameter independent} boundary 
Boltzmann weights. These boundary Boltzmann weights are used
to construct a $Z_N$ model with the fixed-spins along its boundaries.
The star-triangle relation is enough to hold the integrability of the
model with such boundary condition. 
In sections~5 and 6 we consider
the Fateev-Zamolodchikov model and Kashiwara-Miwa model. Apart 
from the fixed-spin boundary conditions, we find another spectral-dependent
boundary Boltzmann weights for the Fateev-Zamolodchikov model. 
In sections~7 the bulk and surface free energies of these two models
are calculated. In particular,  the surface critical exponent is
extracted from the surface free energy deviated from criticality for
the off-critical Kashiwara-Miwa model. Finally a brief
discussion concludes the paper.

\subsection{Integrability}\setcounter{equation}{0}
The non-chiral $Z_N$ model is generally represented by
a planar lattice model with $N$-spins that live on the sites
of the lattice and interact along edges. The Boltzmann weights 
are dependent on a spectral parameter $u$ and a crossing parameter
$\lambda$.  Graphically they can be represented \vspace{-0.5cm}by
\be
W(u|a,b)=\makebox(0,25)[lt]{\face ab{\;\;W(u)}{}}\hs{0.5} \hs{1}
\overline{W}(u|a,b)=\makebox(0,25)[lt]{\face
ab{\;\;\overline{W}(u)}{}} \hs{1}, \label{w}
\ee\vskip 0.6mm\noindent
while the boundary Boltzmann weights are represented by 
\be
K_r(u;\xi_r|a,b)=\makebox(0,25)[lt]{\bond  ab{K(u;\xi_r)}{0}}\hs{1.5} 
K_l(u;\xi_l|a,b)=\makebox(0,25)[lt]{\bond  ab{K(\lambda-u;\xi_l)}{0}}
\hs{2}.\label{k}\ee

\noindent
The spins $a,b=1,2,\cdots,N$. The left $K_l$ and right $K_r$ boundary 
weights may depend on
an arbitrary parameter $\xi_l$ and $\xi_r$ respectively.

The Boltzmann weights share the no-chiral symmetry, or they
are invariant under interchanging spins $a$ and $b$ in 
(\ref{w})-(\ref{k}). We also suppose that they satisfy  
the crossing symmetry
\be
W(u|a,b)=\overline{W}(\lambda-u|a,b)
\ee
and inversion relations
\be
\sum_{d}\overline{W}(u|a,d)\overline{W}(-u|d,c)&=&
\setlength{\unitlength}{0.01050000in}%
\begin{picture}(85,25)(55,797)
\put(100,800){\circle*{5}}
\put( 60,800){\line(1,0){ 80}}
\put(140,790){\makebox(0,0)[lb]{\tiny$c$}}
\put( 55,790){\tiny$a$}
\put(115,805){\tiny$\overline{W}(-u)$}
\put( 70,805){\tiny$\overline{W}(u)$}
\end{picture}\;
=\overline{\rho}(u)\delta_{a,c}   \label{inv1}  \\
\sum_{d}{W}(\lambda-u|a,d){W}(\lambda+u|d,c)&=&
\setlength{\unitlength}{0.01050000in}%
\begin{picture}(85,25)(55,797)
\put(100,800){\circle*{5}}
\put( 60,800){\line(1,0){ 80}}
\put(140,790){\makebox(0,0)[lb]{\tiny$c$}}
\put( 55,790){\tiny$a$}
\put(107,805){\tiny${W}(\lambda\!-\!u)$}
\put(61,805){\tiny${W}(\lambda\!+\!u)$}
\end{picture}\;
={\rho}(u)\delta_{a,c}   \label{inv}\\
{W}(-u|a,d){W}(u|a,d)&=&
\setlength{\unitlength}{0.01050000in}%
\begin{picture}(85,25)(55,797)
\put( 60,800){\line(1,0){ 80}}
\put(140,790){\makebox(0,0)[lb]{\tiny$a$}}
\put( 55,790){\tiny$a$}\put(100,790){\tiny$d$}
\put(100,800){\circle{4}}
\put(107,805){\tiny${W}(-\!u)$}
\put(61,805){\tiny${W}(u)$}
\end{picture}\;
={g}(u){g}(-u)   \label{inv2}\ee
where the solid circles mean sum over all possible spins.
Second inversion relation can be given from  first one
using the crossing symmetry. Thus $\rho(u)=\overline{\rho}(u)$.

Introduce two matrices with a single row of 
alternative Boltzmann weights $W$ and $\overline{W}$ such as
\be
{\tilde V}^{\sigma_1^\prime \sigma_{\!L\!+\!1}^\prime}
       (u)_{\psi,\phi^\prime}&=&
 \prod_{j=1}^{L}\overline{W}(u|\sigma_{j}^\prime,\sigma_{j})
        {W}(u|\sigma_{j},\sigma_{j+1}^\prime)\no\\
&=&\setlength{\unitlength}{0.00550000in}%
\begin{picture}(320,40)(60,740)
\put( 60,780){\line( 1,-1){ 40}}
\put(100,740){\line( 1, 1){ 40}}
\put(140,780){\line( 1,-1){ 40}}
\put(180,740){\line( 1, 1){ 40}}
\put(220,780){\line( 1,-1){ 40}}
\put(260,740){\line( 1, 1){ 40}}
\put(300,780){\line( 1,-1){ 40}}
\put(340,740){\line( 1, 1){ 40}}
\put( 80,760){\tiny$\overline{W}$}
\put(125,760){\tiny$W$}
\end{picture}\\
V^{\sigma_1 \sigma_{\!L\!+\!1}}
                (u)_{\phi,\psi^\prime}&=&\prod_{j=1}^{L}
  \overline{W}(u|\sigma_{j},\sigma_{j}^\prime)
        {W}(u|\sigma_{j}^\prime,\sigma_{j+1})\no\\
&= &\setlength{\unitlength}{0.00550000in}%
\begin{picture}(320,40)(60,740)
\put( 60,740){\line( 1, 1){ 40}}
\put(100,780){\line( 1,-1){ 40}}
\put(140,740){\line( 1, 1){ 40}}
\put(180,780){\line( 1,-1){ 40}}
\put(220,740){\line( 1, 1){ 40}}
\put(260,780){\line( 1,-1){ 40}}
\put(300,740){\line( 1, 1){ 40}}
\put(340,780){\line( 1,-1){ 40}}
\put( 80,760){\tiny$\overline{W}$}
\put(120,760){\tiny$W$}
\end{picture}
\ee
where spin sets 
$\phi=(\sigma_1,\sigma_2,\cdots,\sigma_{L},\sigma_{L+1})$ and
$\psi=(\sigma_1,\sigma_2,\cdots,\sigma_{L})$. 
Then the transfer matrix $T(u)$ is defined by the following 
elements (see fig.\ref{f1})
\begin{figure}[t]
\begin{center}
\setlength{\unitlength}{0.00650000in}%
\begin{picture}(310,45)(0,0)
\put(0,10){\trans {\overline{W}}{\overline{W}}WW0{}{}{}{} }
\put(-17,-7){\bottom}
\put(0,10){\bond {}{}{K_l\!(\!u\!)}0}
\put(320,10){\bond {}{}{K_r\!(\!u\!)}0}
\end{picture}
\caption{\label{f1}\small The transfer matrix $T(u)$:
two rows of sites of the diagonal square lattice with
the boundary weights $K_l(u;\xi_l)$ and $K_r(u;\xi_r)$,
  where the solid circles mean sum.}\end{center}
\end{figure}
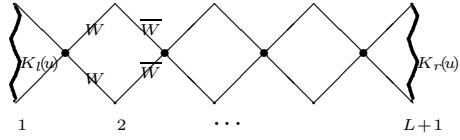
\be
T(u)_{\phi,\phi^\prime}&=&K_l(u|a,a^\prime) V^{ab}(u){\tilde 
    V}^{a^\prime b^\prime}(u)K_r(u|b,b^\prime). \label{2row}   
\ee
The transfer matrix is called as the double-row transfer matrix.
Based on it the planar square lattice can be constructed.
The boundaries on left and right sides of the lattice are not 
periodic. For such lattice the partition function is given by
\be
Z(u)=Tr \biggl( T(u)\biggr)^{M},
\ee
where $M$ is the number of the double-rows in the lattice.

The non-periodic boundary model is integrable if the transfer matrix
forms commuting family
\be
T(u)T(v)=T(v)T(u).
\ee
Suppose that the  bulk Boltzmann weights satisfy
the star-triangle relation
\be\star u{\!v\!\!+\!\!u}v{} \hs{1.5},\label{star}\ee\vskip -0.7cm 
\noindent
where $\chi$ is a spin-independent parameter. 
The star-triangle relation is essential to guarantee the integrability
of a model with a periodic boundary condition.
For the boundary lattice model with the transfer matrix (\ref{2row})
the integrability requires both the
star-triangle relation (\ref{star}) for the bulk weights and 
the following boundary star-triangle relations
\be
\st abc{\!\!W\!(\!u\!\!-\!\!v\!)}{\hs{-0.2}
  W\!(\!u\!\!+\!\!\!v\!)}{K(u)}{K(v)} \label{ref}
\ee\vskip -0.6cm \noindent
for the boundary weights. Given the bulk Boltzmann weights which solve 
(\ref{star}), the boundary Boltzmann weights are determined by solving
(\ref{ref}).

\subsection{Functional relations and fusion hierarchies}
\setcounter{equation}{0}
Fusion procedure is the idea to build the new Boltzmann weights 
using the elementary Boltzmann weights \cite{KuReSk:81}.  The fused transfer
matrices can be constructed with the new Boltzmann weights.
Of particular, the fused transfer matrices form the fusion 
hierarchies  and satisfy a group of functional 
relations. To apply the fusion to the $Z_N$ model let us first
consider  the product of two transfer matrices 
$T(u)T(u+\lambda)$. Using a graph it  can be represented by 
\be
\setlength{\unitlength}{0.00650000in}%
\begin{picture}(110,165)(40,-60)
\put(323,0){\bond  {}{}{K(u\!\!+\!\!\lambda;\xi_r)}{0}}
\put(3,0){\bond{}{}{\hs{-1.2}K(-u;\xi_l)}{0}}
\put(323,-80){\bond  {}{}{K(u;\xi_r)}{0}}
\put(3,-80){\bond{}{}{\hs{-1.3}K(\lambda\!\!-\!\!u;\xi_l)}{0}}
\multiput(322,5)(-320,0){2}{\circle*{6}}
\put(0,0){\trans {\overline{W}}{\overline{W}}{W}{
   W}{6}{a^\prime}{}{b^\prime}{} }
\put(0,-80){\trans 
   {\overline{W}}{\overline{W}}WW0{}a{}b }
\end{picture}
\ee
where the solid circles indicate sum over all possible
spins. Inserting the inversion 
relation (\ref{inv}) into this and using the
star-triangle relation, the product becomes 
\be
\setlength{\unitlength}{0.00650000in}%
\begin{picture}(110,150)(60,-60)
\put(383,0){\bond  {}{}{K(u\!\!+\!\!\lambda;\xi_r)}{0}}
\put(-60,0){\bond{}{}{\hs{-1.2}K(-u;\xi_l)}{0}}
\put(383,-80){\bond  {}{}{K(u;\xi_r)}{0}}
\put(-60,-80){\bond{}{}{\hs{-1.3}K(\lambda\!\!-\!\!u;\xi_l)}{0}}
\multiput(382,5)(-443,0){2}{\circle*{6}}
\put(-10,-5){\tiny$c$}
\put(0,0){\trans 
 {\overline{W}}{W}{W}{\overline{W}}{6}{a^\prime}{}{b^\prime}{} }
\put(0,-80){\trans {W}{\overline{W}}{\overline{W}}{W}0{}a{}b }
\put(0,5){\line(-1,0){60}}\put(0,5){\circle*{6}}
\put(320,5){\line(1,0){60}}\put(320,5){\circle*{6}}
\put(-67,5){\tiny$$} \put(382,1){\tiny$$} 
\put(-53,12){\tiny$W(\lambda\!\!-\!\!2u)$}
\put(322,12){\tiny$W(\lambda\!\!+\!\!2u)$}
\put(-190,0){$\displaystyle1\over\rho(2u)$}
\end{picture}\label{level2}
\ee
This can be divided into two terms according to
the summation over $c$. If $c=a^\prime$, using the inversion
relations (\ref{inv1})-(\ref{inv2}), namely the following
properties
\be\begin{picture}(0,30)(100,15)
\put(0,0){\setlength{\unitlength}{0.008000in}%
\begin{picture}(0,70)(40,0)
\put(0,0){\bwa{a}{c}{\hs{0.1}{W}(u)}{\ig}}
\put(0,30){\wb{}{a}{\overline{W}(u\!+\!\lambda)}{\ig}}
\put(80,35){$=\;g(u){g}(-u)$}
\end{picture}}
\put(180,0){
\setlength{\unitlength}{0.008000in}%
\begin{picture}(0,70)(40,0)
\put(0,0){\wb{a}{}{\hs{0.1}\overline{W}(u)}{\ig}}
\put(0,30){\bwa{}{b}{{W}(u\!+\!\lambda)}{\ig}}
\put(15,39){\circle*{6}}
\put(85,35){$=\;\overline{\rho}(u)\delta_{a,b}$}
\end{picture}}\end{picture}
\ee
for the bulk weights and
\be
\setlength{\unitlength}{0.00450000in}%
\begin{picture}(0,150)(400,-60)
\put(383,-21){\bond  {}{}{K(u\!\!+\!\!\lambda;\xi)}{0}}
\put(383,-80){\bond  {}{}{K(u;\xi)}{0}}
\put(378.5,5){\circle*{8}}
\put(300,5){\line(1,0){80}}
\put(290,0){\small$a$} \put(382,93){\small$a$} \put(382,-87){\small$b$} 
\put(309,12){\tiny$W(\lambda\!\!+\!\!2u)$}
\put(510,0){$=\;\rho_s^{(a)}(u;\xi)\delta_{a,b}$}
\end{picture}\label{rho-s}
\ee
for the boundaries, we are able to show this product gives 
the identity matrix multiplied by a function $f(u)$.
The rest terms with $c\not=a^\prime$ is collected together and
is written as the fused transfer matrix $T^{(2)}(u)$. Thus we have
\begin{equation}
T(u)T(u+\lambda)=f(u)+T^{(2)}(u),\label{T2}
\end{equation}
where 
\be
f(u)&=&(b(u))^2s(u) \label{f}\\
b(u)&=&[g(u){g}(-u)\overline{\rho}(u)]^{L}\label{b(u)}\\
s(u)&=&\rho_s^{(a)}(u;\xi_r)\rho_s^{(a)}(-u;\xi_l).\label{s1}\ee
The transfer matrix $T^{(2)}(u)$ is given by the following 
local fused weights
\be
\begin{picture}(0,35)(110,10)
\put(0,0){
\setlength{\unitlength}{0.008000in}%
\begin{picture}(0,70)(40,0)
\put(0,0){\bwa{a}{c}{\hs{0.1}{W}(u)}{\ig}}
\put(0,30){\wb{}{b}{\overline{W}(u\!+\!\lambda)}{\ig}}
\end{picture}}
\put(80,0){
\setlength{\unitlength}{0.008000in}%
\begin{picture}(0,70)(40,0)
\put(0,0){\wb{a}{}{\hs{0.1}\overline{W}(u)}{\ig}}
\put(0,30){\bwa{c}{b}{{W}(u\!+\!\lambda)}{\ig}}
\put(15,39){}
\end{picture}}
\put(140,20){and}
\put(230,0){\setlength{\unitlength}{0.00450000in}%
\begin{picture}(0,150)(400,-60)
\put(383,-21){\bond  {}{}{K(u\!\!+\!\!\lambda;\xi)}{0}}
\put(383,-80){\bond  {}{}{K(u;\xi)}{0}}
\put(378.5,5){\circle*{8}}
\put(300,5){\line(1,0){80}}
\put(290,0){\small$a$} \put(382,93){\small$d$} \put(382,-87){\small$b$} 
\put(309,12){\tiny$W(\lambda\!\!+\!\!2u)$}
\end{picture}}\end{picture}
\ee
where $a\ne b$, $a\ne d$ and $c\ne N$. For clarity $T^{(2)}(u)$
can be represented by a similar graph in  equation (\ref{level2}).

Consider $n$ by $m$ square lattice with the spectral parameters
shifted properly \cite{ZhPe:94}. We can fuse
it to obtain the fused weights and thus define the fused
transfer matrices $T^{(m,n)}(u)$, where two integers $m,n$ 
greater than $1$ are the fusion levels . For example, the above 
fused transfer matrix given in (\ref{T2}) 
corresponds to $T^{(1,2)}(u)$. Note that 
the fused boundary weights are labelled with a single fusion 
level $n$. Here we are not interested in the detail of showing 
the procedure of fusion. Instead, we write down the functional 
relations of the fused transfer matrices. 
Generalising (\ref{T2}), we are able to derive 
\be
T^{(m,n)}(u)T^{(m,1)}(u+n\lambda)=T^{(m,n+1)}(u)  + 
  f^{(m)}_{n-1}T^{(m,n-1)}(u),  \label{fun} \ee
where $T^{(m,0)}(u)=I$ and 
\be
f^{(m)}_{n}&=&f^{(m)}(u+n\lambda) \label{fm} \\
f^{(m)}(u)&=&s(u)b^{(m)}(u)  \label{fm1} \\
b^{(m)}(u)&=&  \prod_{j=0}^{m-1}b(u-j\lambda)b(u+j\lambda)
\ee
The above functional relations imply
\be
T^{(m,n)}(u)T^{(m,n)}(u+\lambda)&=&\prod_{k=0}^{n-1}f^{(m)}_k  
   + T^{(m,n+1)}(u)T^{(m,n-1)}(u+\lambda).
\label{TT}\ee
The functional relations (\ref{fun}) and (\ref{TT}) are the
$T$-system of the model. From them the $y$-system can be introduced.
Define
\be
y^{(m,0)}(u)&=&0 \;     \\
y^{(m,n)}(u)&=&\displaystyle{T^{(m,n+1)}(u)T^{(m,n-1)}(u+\lambda)/ 
     \prod_{k=0}^{n-1}f^{(m)}_k }\; ,
\label{def-y}
\ee
From (\ref{TT}) we are  able to derive  the $y$-system 
\be
y^{(m,n)}(u)y^{(m,n)}(u+\lambda)=(1+y^{(m,n+1)}(u))
 (1+y^{(m,n-1)}(u+\lambda))\;.\label{func-y}
\ee
These functional relations are $su(2)$ type. They 
have been seen for other models with periodic boundaries
\cite{KiRe:87,BaRe:89,KlPe:92,ZhPe:94,Zhou:96b} and reflection
boundaries \cite{Zhou:95,Zhou:96a,BOP:95,ZhBa:96,Zhou:96c}.

\subsection{Fixed-spin boundaries}\setcounter{equation}{0}
\begin{figure}[t]
\begin{center}
\setlength{\unitlength}{0.00650000in}%
\begin{picture}(310,45)(0,0)
\put(0,10){\trans{\overline{W}}{\overline{W}
  }WW0{\alpha}{\alpha^\prime}{\beta}{\beta^\prime} }
\put(-17,-7){\bottom}
\end{picture}
\caption{\small \label{fig2} 
 Two rows of sites of the  square lattice with
the boundary spins being fixed, which defines the transfer matrix 
$T_{(\alpha\alpha^\prime,\beta\beta^\prime)}(u)$.}
\end{center}
\end{figure}
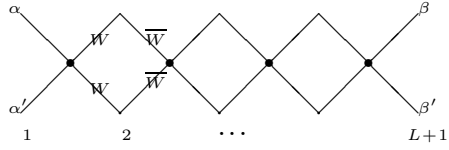
A common future of the $Z_N$  model is the existence of
the following boundary weights
\be
K(u|c,d)=\delta_{a,\alpha}\delta_{b,\beta}\label{k1}
\ee
for the given boundary spins $\alpha,\beta$.
Obviously, this $K$-matrix solves the boundary star-triangle relations 
(\ref{ref}). With this boundary weights the 
square lattice with fixed boundary spins can be constructed. 
The transfer matrix is given by  Fig~\ref{fig2}.

The functional relations given in the last section are still valid for
the fixed boundary lattice. But the function $f(u)$ is changed. To see
this let us consider the product of
$T_{(\alpha\alpha^\prime,\beta\beta^\prime)}(u)
T_{(\alpha^\prime\alpha,\beta^\prime\beta)}(u+\lambda)$ and this yields
functional relation (\ref{T2}). The involved functions
in the functional relations (\ref{T2}), (\ref{fun}) and
(\ref{TT})-(\ref{func-y}) are given by (\ref{f}) and (\ref{fm}) 
with
\be
s(u)=W(\lambda-2u|\alpha,\alpha^\prime)
W(\lambda+2u|\beta,\beta^\prime),\label{s2}\ee
while $b(u)$ and $b^{(m)}(u)$ remain unchanged.

\subsection{Fateev-Zamolodchikov model}\setcounter{equation}{0}
From this section we start to consider the specific non-chiral
models. We apply the general formulae given previously
to the Fateev-Zamolodchikov model in this section and to
the Kashiwara-Miwa model in next section.

The Fateev and Zamolodchikov is described by the following
bulk Boltzmann weights $W(u|a,b)$ and 
$\overline{W}(u|a,b)$  \cite{FaZa:82}
\be
{W(u|a,b)}&=&g(u)\prod_{j=1}^{|a-b|}{\sin((2j-1)\lambda-u)
   \over\sin((2j-1)\lambda+u)}   \\
{\overline{W}(u|a,b)}&=&\overline{g}(u)\prod_{j=1}^{|a-b|}{
  \sin(2(j-1)\lambda+u)
   \over\sin(2j\lambda-u)}  \label{FZ-W}
\ee
with
\be
g(u)&=&\prod_{j=1}^{n}\sin((2j-1)\lambda+u)  \label{g}\\
\overline{g}(u)&=&\prod_{j=1}^{n}\sin(2j\lambda-u)
\ee
where  we use $n=(N-1)/2$ for odd $N$, $n=N/2$ 
for even $N$ and  $\lambda={\pi/ 2N}$. 
For this model the parameter $\chi=\sqrt{N}$
in the star-triangle relation and 
$\overline{\rho}(u)=g(u)g(-u)\overline{g}(0)(g(0)\chi)^{-1}$ 
in the inversion relation (\ref{inv1}).

The boundary Boltzmann weights are given by solving the boundary
star-triangle relations (\ref{ref}). Apart from the fixed-spin boundary 
solution (\ref{k1}), we also present the following solution
\be
K(u;\xi|a,b)=K(u;\xi|a-b)=K(u;\xi|b-a)=0, 
 \hs{1}\mbox{unless $|a-b|=0$ or $N-1$}
\ee
and 
\be
K(u;\xi|N-1)={C K(u;\xi|0)\sin(2u)} \hs{0.5}\mbox{and}\hs{0.5}
K(u;\xi|0)={\rm arbitrary}
\ee
where $C$ is $u$-independent and depends on $\xi$. 
The above solution implies that the function in (\ref{rho-s}) 
\be\rho^{(a)}_s(u;\xi)=g(\lambda+2u)K(u|0)K(\lambda+u|0)(
1-C^2\epsilon \sin^2(2u))\label{el}\ee
and $s(u)$ is given by (\ref{s1}) for the Fateev-Zamolodchikov model,
where $\epsilon=1$ for $a=1,N$ and $\epsilon=0$ otherwise.

\subsection{Kashiwara-Miwa  model}\setcounter{equation}{0}
The Kashiwara-Miwa model is the elliptic extension of
the Fateev-Zamolodchikov model. The bulk Boltzmann weights 
are given by  \cite{KaMi:86}
\be
{W(u|a,b)}&=&
 e^{-ug_{a,b}}g(u)\prod_{j=1}^{|a-b|}{\vartheta_1((2j-1)\lambda-u)
   \over\vartheta_1((2j-1)\lambda+u)}
  \prod_{j=1}^{a+b+s}{\vartheta_4((2j-1)\lambda-u)
   \over\vartheta_4((2j-1)\lambda+u)}  \label{kmw1} \\
{\overline{W}(u|a,b)}&=&e^{(u-\lambda)g_{a,b}}
   \overline{g}(u)\prod_{j=1}^{|a-b|}{
  \vartheta_1(2(j-1)\lambda+u) \over\vartheta_1(2j\lambda-u)}
   \prod_{j=1}^{a+b+s}{\vartheta_4(2(j-1)\lambda+u)
   \over\vartheta_4(2j\lambda-u)} \label{kmw2}
\ee
with
\be
g(u)&=&\prod_{j=1}^{n}\vartheta_1((2j-1)\lambda+u)
  \vartheta_4((2j-1)\lambda+u) \label{kg} \\
\overline{g}(u)&=&\prod_{j=1}^{n}
  \vartheta_1(2j\lambda-u)\vartheta_4(2j\lambda-u)
\ee
where $g_{a,b}=\log(R_a R_b)/\lambda $ and
$R_a=\sqrt{\vartheta_4(2s\lambda)/\vartheta_4(4a\lambda+2s\lambda)}$.
$s=0$ or $1$ for even $N$ and $s=0$ for odd $N$. The elliptic 
functions $\vartheta_1({u})$, $\vartheta_4({u})$ are standard 
theta functions of nome $p$ 
\be
\vartheta_1(u)&=&\vartheta_1(u,p)=2p^{1/4}\sin u\:
  \prod_{k=1}^{\infty} \left(1-2p^{2k}\cos
   2u+p^{4k}\right)\left(1-p^{2k}\right)\label{theta1}  \\
\vartheta_4(u)&=&\vartheta_4(u,p)=\prod_{k=1}^{\infty}\left(
 1-2p^{2k-1}\cos2u+p^{4k-2}\right)\left(1-p^{2k}\right)
  \label{theta4}
\ee
where $0<p<1$ with $p=0$ at criticality. The parameter 
$\chi=\overline{W}(0|0,0)/\overline{W}(\lambda|0,0)$
in the star-triangle relation and 
$\overline{\rho}(u)=g(u)g(-u)\overline{W}(\lambda|0,0)/{W}(0|0,0)$ 
in the inversion relation (\ref{inv1}).

The boundary Boltzmann weights are given by (\ref{k1}).
Unfortunately, we have not found other solutions 
for the Kashiwara-Miwa model. For the
fixed-spin boundary the function $s(u)$ given in (\ref{s2})
is still valid.

\subsection{Surface free energies}\setcounter{equation}{0}
The surface free energies of the $Z_N$ models together with their 
bulk free energies can be found from the functional relation
(\ref{T2}) as the other models shown in  
\cite{Zhou:96c,Zhou:96a}. In $L\to\infty$ the 
fused transfer matrix $T^{(2)}(u)$ stands for the finite-size 
corrections of the transfer matrix $T(u)$. Therefore the bulk 
and surface free energies  satisfy
\be
{T}(u){T}(u+\lambda)=f(u)\;, \label{unit}
\ee
The unitarity relation (\ref{unit})  
combines the inversion relation and crossing symmetries of 
the local bulk and boundary face weights. We can separate
the bulk free energy from the surface free energy 
\cite{Zhou:96a,Zhou:96b}. Let $T(u)=T_b(u) T_s(u)$ be the 
eigenvalues of the transfer matrix $\mbox{\boldmath $T$}(u)$. Define 
$T_b = \kappa_b^{2 L}$ 
and $T_s = \kappa_s$, then the bulk and surface
free energies are defined by $f_b(u)=-\log \kappa_b(u)$ 
and $f_s(u) = -\log \kappa_s(u)$ respectively.
We have
\be
\log T(u)=-2L f_b(u)-f_s(u)+... \hs{0.5}{\rm as}\hs{0.5}L\to\infty.
\ee
The terms are represented by the dots are the finite-size
corresctions. 

In present paper we calculate the free energies
$f_b(u)$ and $f_s(u)$ for the Fateev-Zamolodchikov model
and Kashiwara-Miwa model. Seperating the surface free
energy from the bulk one in (\ref{unit}), we are
able to obtain
\be
\kappa_b(u)\kappa_b(u+\lambda)&=&g(u)g(-u)\overline{\rho}(u)  \label{kb}\\
\kappa_s(u)\kappa_s(u+\lambda)&=&s(u) .\label{ks}
\ee
These inversion relations can be solved  with certain analyticity 
assumptions. This is known as ``inversion relation trick''. The trick  
has been successfully applied to give the correct
bulk free energy  \cite{Baxter:82b} and
surface free energy \cite{BaZh:95} of the eight-vertex model. 

\subsubsection{Fateev-Zamolodchikov model}
The bulk free energy of the Fateev-Zamolodchikov model has been given
using the inversion relation trick in \cite{FaZa:82} and
the  root density method in \cite{Albertini:92}. 
Here the normalisation of the bulk Boltzmann weights (\ref{FZ-W}) 
is different from that  adopted in \cite{FaZa:82,Albertini:92},
or $W(u|a,a)\not=1$ and $\overline{W}(u|a,a)\not=1$. So it is 
worthwhile exercise to derive the bulk free energy before
calculating the surface free energy.

\noindent{7.1.1 \hs{0.2}Bulk free energy}\smallskip\\
The bulk inversion relation (\ref{kb}) becomes
\be
\kappa_b(u)\kappa_b(u+\lambda)=\!\prod_{j=1}^{n}{\sin((2j-1)\lambda+u)
  \sin((2j-1)\lambda-u)\over\sin^2((2j-1)\lambda) }\label{fzkb}
\ee
for the Fateev-Zamolodchikov model. To solve it let us
suppose that $\kappa(u)$ is analytic and non-zero in 
the regime $0<u<\lambda$.
 Changing variable $u=ix$ and applying the
following Fourier transforms
\be 
&& K(k)={1\over 2\pi}\int_{-\infty}^{\infty}dx\;
     [\ln\kappa(x)]^{\prime\prime}\;e^{i kx},\no\\ 
&&\hs{0.5} [\ln\kappa(x)]^{\prime\prime}=\int_{-\infty}^{\infty} 
  dk\;K(k) \;e^{-i kx} \; \label{FT}
\ee 
to (\ref{fzkb}), then solving the equation for $K(k)$ and inverting
Fourier transform back to $\kappa(x)$, we have
\be
[\ln\kappa(x)]^{\prime\prime}=\int_{-\infty}^{\infty}dk\;k\;
 e^{(-ix+\lambda/2)k}\sum_{j=1}^n
 {\cosh[((2j-1)\lambda-\pi/2)k]\over \sinh(k\pi/2)\cosh(k\lambda/2)}.
\ee
Integrating twice and using the condition 
$\kappa(u)=1$ as $u=0$ and $u=\lambda$,
we obtain the bulk free energy
\be
f_b(u)=-2\int_{-\infty}^\infty {dk}F_b(k,u),
\ee
where 
\be
F_b(k,u)&=&{\sinh(ku)\sinh(k\lambda\!-\!ku)\over 
        k\sinh(k\pi)\cosh(k\lambda)}
 {\sinh(2kn\lambda)\cosh(2kn\lambda\!-\!k\pi)
    \over\sinh(2k\lambda)}. \label{Fb}\ee

\noindent{7.1.2\hs{0.2} Surface free energies}\smallskip\\
The surface free energies are consist of the excess and local surface 
free energies \cite{ws,bin,diehl}, 
or $f_s(u)=f_s^s(u)+f_s^l(u)$.  Accordingly, the inversion 
relation (\ref{ks}) for the Fateev-Zamolodchikov model
breaks into two parts determining the excess and local surface 
free energies respectively. Suppose 
$$\kappa_s(u)=\kappa_s^s(u)\kappa_s^l(u).$$
Then the  excess free energy reads 
$$f_s^s(u)=-\log\kappa_s^s(u).$$ 
and the local free energy reads 
$$f_s^l(u)=-\log\kappa_s^l(u).$$
Their inversion relations are given by
\be
\kappa_s^s(u)\kappa_s^s(u+\lambda)&=&\prod_{j=1}^{n}{\sin(2j\lambda+u)
  \sin(2j\lambda-u)\over\sin^2(2j\lambda) }\label{kss}\\
\kappa_s^l(u)\kappa_s^l(u+\lambda)
  &=&K(u|0)K(\lambda+u|0)K(-u|0)K(\lambda-u|0)
\no\\&& \times(1-C^2_l \sin^2(2u))(1-C^2_r\sin^2(2u)), \label{ksl}
\ee 
where the local free energy occurs only for $a=1,N$ in (\ref{el}), or
the boundary couplings appear.
Furthermore, we can divide the local free energy into the
left  and right free energies, 
\be
f_s^l(u)=f_s^{ll}(u)+f_s^{lr}(u) \\
\kappa_s^{l}(u)=\kappa_s^{ll}(u)\kappa_s^{lr}(u).
\ee
Taking 
\be
C_l=1/\sin{\xi^l} \hs{0.5}C_r=1/\sin{\xi^r}
\ee
and the proper normalisation $K(u|0)$, we are able to obtain
\be
\kappa_s^{ll}(u)\kappa_s^{ll}(u+\lambda)&=&
   {\sin(\xi^l-2u)\sin(\xi^l+2u)\over\sin^2\xi^l} \label{kll}\\
\kappa_s^{lr}(u)\kappa_s^{lr}(u+\lambda)&=&
  {\sin(\xi^r-2u)\sin(\xi^r+2u)\over\sin^2\xi^r}.\label{klr}
\ee

As the bulk free energy, we suppose that 
$\kappa_s^s(u)$, $\kappa_s^{ll}(u)$ and $\kappa_s^{lr}(u)$  
are analytic and non-zero in the regime $0<u<\lambda$.
Using the Fourier transforms (\ref{FT}), the inversion relations
(\ref{kss}) and (\ref{kll})-(\ref{klr}) 
can be  solved similarly.  Define two functions
\be
F_s(k,u)={\sinh(ku)\sinh(k\lambda\!-\!ku)
    \over k\sinh(k\pi/2)\cosh(k\lambda)}
 {\sinh(kn\lambda)\cosh(k\lambda(n\!+\!1)\!-\!k\pi/2)
   \over\sinh(k\lambda)} \label{Fs}\\
F_l(k,u,\xi)=  {\sinh(ku)\sinh(k\lambda\!-\!ku)
  \cosh(k\xi\!-\!k\pi/2)
  \over k\cosh(k\lambda)\sinh(k\pi/2)}.\label{Fl}
\ee
We are able to obtain the excess surface 
free energy
\be
f_s(u)=-2\int_{-\infty}^\infty {dk}F_s(k,u),
 \ee
the left local surface free energy
\be
f_s^{ll}(u)=-2\int_{-\infty}^\infty {dk}F_l(k,u,\xi^l)
\ee
and the right local surface free energy
\be
f_s^{lr}(u)=-2\int_{-\infty}^\infty {dk}F_l(k,u,\xi^r).
\ee
To derive the local free energies the surface coupling parameters 
have been restricted as $0<\xi^{l,r}<\pi$.

\subsubsection{Kashiwara-Miwa model}
The function $s(u)$ and $b(u)$ are given by (\ref{s2}) and (\ref{b(u)})
respectively for the Kashiwara-Miwa
model with fixed-spin boundaries. For simplicity,
we take the boundary spins $\alpha^\prime=\alpha$ and $s=0$ in the 
bulk Boltzmann weights (\ref{kmw1})-(\ref{kmw2}). We can write
the inversion relations (\ref{kb})-(\ref{ks}) as
\be
\kappa_b(u)\kappa_b(u+\lambda)=\!\prod_{j=1}^{n}\!
  {\vartheta_1((2j\!\!-\!\!1)\lambda\!\!+\!\!u)
   \vartheta_4((2j\!\!-\!\!1)\lambda\!\!+\!\!u)
   \vartheta_1((2j\!\!-\!\!1)\lambda\!\!-\!\!u)
   \vartheta_4((2j\!\!-\!\!1)\lambda\!\!-\!\!u) \over 
  \vartheta_1((2j-1)\lambda)
  \vartheta_4((2j-1)\lambda)\vartheta_1((2j-1)\lambda)
  \vartheta_4((2j-1)\lambda) } \label{kb1}\\
\kappa_s(u)\kappa_s(u\!+\!\lambda)=\prod_{j=1}^{n}
  {\vartheta_1(2j\lambda\!+\!2u)
   \vartheta_4(2j\lambda\!+\!2u)
   \vartheta_1(2j\lambda\!-\!2u)
   \vartheta_4(2j\lambda\!-\!2u) \over 
  \vartheta_1(2j\lambda)
  \vartheta_4(2j\lambda)\vartheta_1((2j-1)\lambda)
  \vartheta_4(2j\lambda) } \label{ks1}
\ee
The bulk free energy has been calculated in \cite{JMO:86}. Here we
have taken the different normalisation. So we present the bulk free
energy again along with the surface free energy.

To solve the unitarity relations (\ref{kb1})-(\ref{ks1})  
let us  introduce the new variables
\be
x=e^{-\pi\lambda/\epsilon}, &
w=e^{-2\pi u/\epsilon},&
q = e^{-\pi^2/\epsilon} \hs{0.4}p=e^{-\epsilon} 
\ee
along with the conjugate modulus transformation of
the theta functions,
\be
\vartheta_1(u,e^{-\epsilon})&=&\rho(u,\epsilon) 
     E\left(e^{-2\pi u/\epsilon},
        e^{-2\pi^2/\epsilon}\right) \\
\vartheta_4(u,e^{-\epsilon})&=&\rho(u,\epsilon) 
     E\left(-e^{-2\pi u/\epsilon},
        e^{-2\pi^2/\epsilon}\right). 
\ee
The factor $\rho(u,\epsilon)$ is harmless and will be 
disregarded, while
\be
E(z,x)=\prod_{n=1}^\infty(1-x^{n-1}z)(1-x^{n} z^{-1})(1-x^n).
\ee

Suppose that $\kappa_b(w)$ is analytic and nonzero 
in the annulus $x\le w\le 1$ and perform the Laurent expansion  
$\log\kappa_b(w)=\displaystyle{\sum_{m=-\infty}^{\infty} c_m w^m}$.
Then inserting this into the logarithm of both sides of (\ref{kb1})
and equating coefficients of powers of $w$ gives 
\be
f_b(w,p)&=&-{2\pi\over\epsilon}\sum_{m=1}^{\infty}(1+(-1)^m)
            F_b(\pi m/\epsilon,u) \no\\
   &=&-{4\pi\over\epsilon}\sum_{m=1}^{\infty}
           F_b(2\pi m/\epsilon,u)\label{bfree}
\ee
Similarly, Laurent expanding $\log\kappa_s(w)=
 \sum_{m=-\infty}^{\infty} c_m w^m$ and solving the
crossing unitarity relation (\ref{ks1}) under the same 
analyticity assumptions as for the bulk case gives the 
excess surface free energy
\be
f_s(w,p)&=&-{4\pi\over\epsilon}\sum_{m=1}^{\infty}(1+(-1)^m)
                    F_s(2\pi m/\epsilon,u) \no\\
   &=&-{8\pi\over\epsilon}\sum_{m=1}^{\infty}
                  F_s(4\pi m/\epsilon,u).\label{sfree}
\ee
The local surface free energy is not obtainable from
the fixed-spin boundaries. Moreover, the excess surface free
energy will be changed if the boundary spin $\alpha^\prime\not=\alpha$.

\subsubsection{Critical properties}
The Kashiwara-Miwa model is more interesting here
as their solution in terms of elliptic functions correspond to
off-critical extensions of the the Fateev-Zamolodchikov model.
The elliptic nome $p$ measures the
deviation from the critical point $p=0$. 
 
The critical behaviour of free energies is described by their
the singular behaviour in $p\to 0$. 
In practice, the singular  part of the free energies given 
in (\ref{bfree}) and (\ref{sfree}) are extracted by means of
the Poisson summation formula \cite{Baxter:82}. For the bulk free 
energy $f_b(w,p)$ it follows that \cite{JMO:86} 
\be
f_b(w,p)\sim p^{N}\log p\quad\mbox{as}\quad p\to 0. \label{bs}
\ee
For the surface free energy a similar treatment yields that 
\begin{eqnarray}
f_s(w,p)\sim \cases{p^{N/2}\log p \quad &$N$ even\cr \quad 
\hs{-0.3}p^{N/2}   & $N$ odd\cr}\quad\mbox{as}\quad p\to 0.\label{ss}
\end{eqnarray}

According to the well developed phenomenology of 
critical behaviour at a surface \cite{bin,diehl,ws}, the excess surface 
critical exponent $\alpha_s$ can be obtained from the surface 
free energy. Together with the bulk critical exponent $\alpha_b$,
which has already been given in \cite{JMO:86},
we have that
\be
f_b(w,p)\sim p^{2-\alpha_b}\log p \quad \mbox{and} \quad 
f_s(w,p)\sim p^{2-\alpha_s} (\log p). \label{al}
\end{eqnarray}
Comparing (\ref{al}) with (\ref{bs}) and (\ref{ss}), we are able to
obtain that
\be
\alpha_b&=&2-N  \\
\alpha_s&=&2-N/2 \label{als}
\ee
and this verifies the scaling relation
\be
\alpha_b=2\alpha_s-2.\label{sc}
\ee

The scaling relation (\ref{sc}) is consistent with the known scaling 
relation $\alpha_s = \alpha_b + \nu$ and $\alpha_b/2=1-\nu$.
Using the known scaling relation $\alpha_s = \alpha_b + \nu$ and 
$\alpha_1=\alpha_b-1$, the results (\ref{als}) imply that
the correlation length critical exponent $\nu=N/2$ and 
the local specific heat critical exponent $\alpha_1=1-N$.

\subsection{Discussion}\setcounter{equation}{0}
In this paper the boundary star-triangle relations and the boundary 
weights for the Fateev-Zamolodchikov and Kashiwara-Miwa
models have been studied. 
Of particular, it has been shown that the fixed-spin boundary
lattices are always integrable provided the bulk Boltzmann weights
satisfy the star-triangle relation. This is even true
for the chiral Potts model. 

The free energies of the Fateev-Zamolodchikov and Kashiwara-Miwa
models have been calculated. The inversion relation determining
the surface free energies is separated from the bulk one
\cite{Zhou:96a}. This leads to that Baxter's inversion trick
\cite{Baxter:82b} is applicable to calculating the surface free 
energy of  integrable models with reflection boundaries. Of course,
the calculation is correct because the analytic assumption is valid to
the models.
The surface critical exponent of the excess specific heat of the
Kashiwara-Miwa model has been given. Using the relevant
known scaling relations, the critical exponents of
the correlation length $\nu$ and the local specific heat $\alpha_1$
are predicted for the Kashiwara-Miwa model.

The study presented in this paper is to be extended to
the chiral Potts model. The boundary Boltzmann weights of the model
have been 
calculated and the further investigation is in  progress.

\subsection*{Acknowledgements} This research has been supported 
by the Australian Research Council and partially by Natural Science
Foundation of China.  


\end{document}